\documentclass{article}
\usepackage{spconf,amsmath,graphicx}
\usepackage{bm,amsfonts}
\usepackage{amsmath,amssymb}    

\usepackage{graphicx}   
\usepackage{verbatim}   
\usepackage{color}      
\usepackage{subfigure}  
\usepackage{hyperref}   
\usepackage{algorithm}\usepackage[noend]{algpseudocode} 

\newcommand{\ds}{{\displaystyle}}

\def\T{{\mathcal T}}
\def\L{{\mathbb L}}

\def\R{{\mathbb R}}

\newcommand{\abs}[1]{\left\lvert#1\right\rvert}
\newcommand{\dotprod}[1]{\langle#1\rangle}
\DeclareMathOperator{\vect}{vec}

\graphicspath{{figs/}}

\title{Low-Rank Tensor Regression for X-Ray Tomography}

\name{Sanket R. Jantre$^{\star}$ \qquad Zichao Wendy Di$^{\dagger}$}
  
  \address{$^{\star}$ Department of Statistics and Probability, Michigan State University, East Lansing, MI, USA\\
      $^{\dagger}$ Mathematics and Computer Science Division, Argonne National Laboratory, Lemont, IL, USA}
	
\begin{document}

%

%

%
\maketitle
\begin{abstract}
Tomographic imaging is useful for revealing the internal structure of a 3D sample. Classical reconstruction methods treat the object of interest as a vector to estimate its value. Such an approach, however, can be inefficient in analyzing high-dimensional data because of the underexploration of the underlying structure. In this work, we propose to apply a tensor-based regression model to perform tomographic reconstruction. Furthermore, we explore the low-rank structure embedded in the corresponding tensor form. As a result, our proposed method efficiently reduces the dimensionality of the unknown parameters, which is particularly beneficial for ill-posed inverse problem suffering from insufficient data. We demonstrate the robustness of our proposed approach on synthetic noise-free data as well as on Gaussian noise-added data. 
\end{abstract}

\begin{keywords}
inverse problem, low-rank approximation
\end{keywords} 



\maketitle


\section{INTRODUCTION} 

\label{Section1} 


Tomographic imaging reconstructs a 3D object volume from its 2D projection images by sectioning through the use of any kind of penetrating wave. This technique is used by numerous fields such as radiology, archaeology, astronomy, and materials science. Various tomogram modalities are derived from diverse physical phenomena \cite{Oldendorf-1978,Mansfield-Grannell-1975,Ter-Pogossian-et-al-1980}. In X-ray tomography (XRT), X-rays are used to visualize the internal structure of an object nondestructively. For example, X-ray absorption provides the spatial distribution of the attenuation coefficient of a 3D object by acquiring a sequence of 2D projections at various angles. The technique  has wide applications in medical imaging, materials science, and geology \cite{Baruchel-2004,Mees-et-al-2003}. The capability of such a technique highly depends on the image reconstruction quality. However, the problem is often ill-posed because of the  limited amount of data and hence does not have a unique solution \cite{Davison-1983,Di-et-al-2016}. 

 The conventional reconstruction algorithms discretize the object region and estimate the unknown property (e.g., attenuation coefficient) for each discretized pixel from various projections. The most commonly used methods to reconstruct the unknown parameters include  analytical reconstruction techniques (e.g., filtered back projection) and various iterative methods (e.g., algebraic reconstruction techniques) \cite{Kak-Slaney-1988}, where the latter are often preferred because of their robustness to data noise and their flexibility in incorporating more accurate imaging models and constraints \cite{Desai-Kulkarni-1989,di2019optimization,di2017joint}. In particular, statistical algorithms (e.g., expectation maximisation  \cite{Dempster-et-al-2000}) try to estimate the solution that maximizes the likelihood of observing the measured projections. A common characteristic of the existing approaches is that the unknown parameters, whose natural form is either a 2D image or a 3D volume, are vectorized first and solved as a 1D vector. Naively turning an image array into a vector is  unsatisfactory, however. For instance, a typical 128-by-128 XRT projection   implicitly requires 16,384 regression parameters---almost always more than the available sample points for estimation. Both computability and theoretical guarantee of the classical regression models are compromised by this high dimensionality. More severe is the fact that vectorizing an array destroys the natural structure and correlation of the image \cite{Tao-2007}.

Alternatively one could consider optimizing the unknown parameters in  original form (e.g., 2D matrix or higher-order tensor) to fully exploit the underlying embedded structure and accelerate the overall estimation efficiency. For example, \cite{Zhou-et-al-2013} proposed a tensor regression model to efficiently estimate the regression coefficients. The main advantage is that by utilizing the CANDECOMP/PARAFAC (CP) decomposition \cite{Kolda-Bader-2009}, this method dramatically reduces the dimensionality of the parametric model while effectively recovering the spatial distribution of the object. Furthermore, this method still retains the flexibility of the traditional regression model in that it allows the incorporation of any prior knowledge in the form of regularization. Regularized tensor regression not only handles the small-sample-large-parameters challenge that is  common in tomography; it  also stabilizes the estimates in well-posed problems. 


In this work, we employ the tensor regression model and apply it for tomographic reconstruction. To overcome the ill-conditioned nature of tomography, we further study the regularized version in order to incorporate prior knowledge. 


\section{MATHEMATICAL FORMULATION} 

\label{Section2} 


Throughout the article, bold lowercase letters $(\bm{l})$ denote vectors, bold uppercase letters $(\bm{L})$ denote matrices, and bold uppercase blackboard letters $(\mathbb{L})$ denote multiway arrays. We first define a few necessary tensor operations following the convention presented in \cite{Kolda-Bader-2009,bader2006algorithm}. Given a $D$-way array (i.e., a $D$th-order tensor) $\mathbb{L}\in \R^{p_1 \times \dots \times p_D}$ with entries denoted as $\mathbb{L}[i_1,\dots,i_D]$, where $i_d \in \{1,\dots,p_d \}$ for $d \in \{1,\dots,D\}$,  we define the \textit{outer product}
$\mathbb{L} = \bm{l}_1 \circ \bm{l}_2 \circ \dots \circ \bm{l}_D,$
where $\bm{l}_1,\dots,\bm{l}_D$ are vectors of length $p_1,\dots,p_D$, respectively. Therefore, we have $\ds \mathbb{L}[i_1,\dots,i_D] = \prod\limits_{d=1}^D \bm{l}_d[i_d]$. Note that the outer product of two vectors is a rank-1 matrix. For matrices $\bm{L}_1,\dots,\bm{L}_D$ having the same column dimension as $R$, we introduce the notation
\begin{equation}\label{CP_decomp}   
    [\![\bm{L}_1,\dots,\bm{L}_D ]\!] = \sum_{r=1}^R \bm{l}_{1}^{(r)} \circ \dots \circ \bm{l}_{D}^{(r)},
\end{equation}
where $\bm{l}_{d}^{(r)}$ is the $r$th column of $\bm{L}_d$. Equation~(\ref{CP_decomp}) defines a CP decomposition, and an array that can be expressed in this form is defined to have rank $R$.

The vectorization operator, $\vect(.)$, transforms a multiway array into a vector following the lexicographical (column-first) order. Specifically, vec($\mathbb{L}$) is a vector of length $\prod\limits_{d=1}^D p_d$. Given $\mathbb{L}\in \R^{I_1\dots\times I_D \times P_1 \dots \times P_A}$ and $\mathbb{W}\in \R^{P_1 \dots\times P_A \times Q_1 \dots \times Q_B}$, we define their \textit{contracted tensor product} 
as
\begin{align*}
    & \dotprod{\mathbb{L},\mathbb{W}}_A [i_1,\dots i_D,q_1,\dots,q_B] = \\
    & \sum_{p_1=1}^{P_1}\dots \sum_{p_A=1}^{P_A} \mathbb{L}[i_1,\dots,i_D,p_1,\dots,p_A] \mathbb{W}[p_1,\dots,p_A,q_1,\dots,q_B].
\end{align*}

\begin{figure}[htb]
\centering
\includegraphics[scale=0.25]{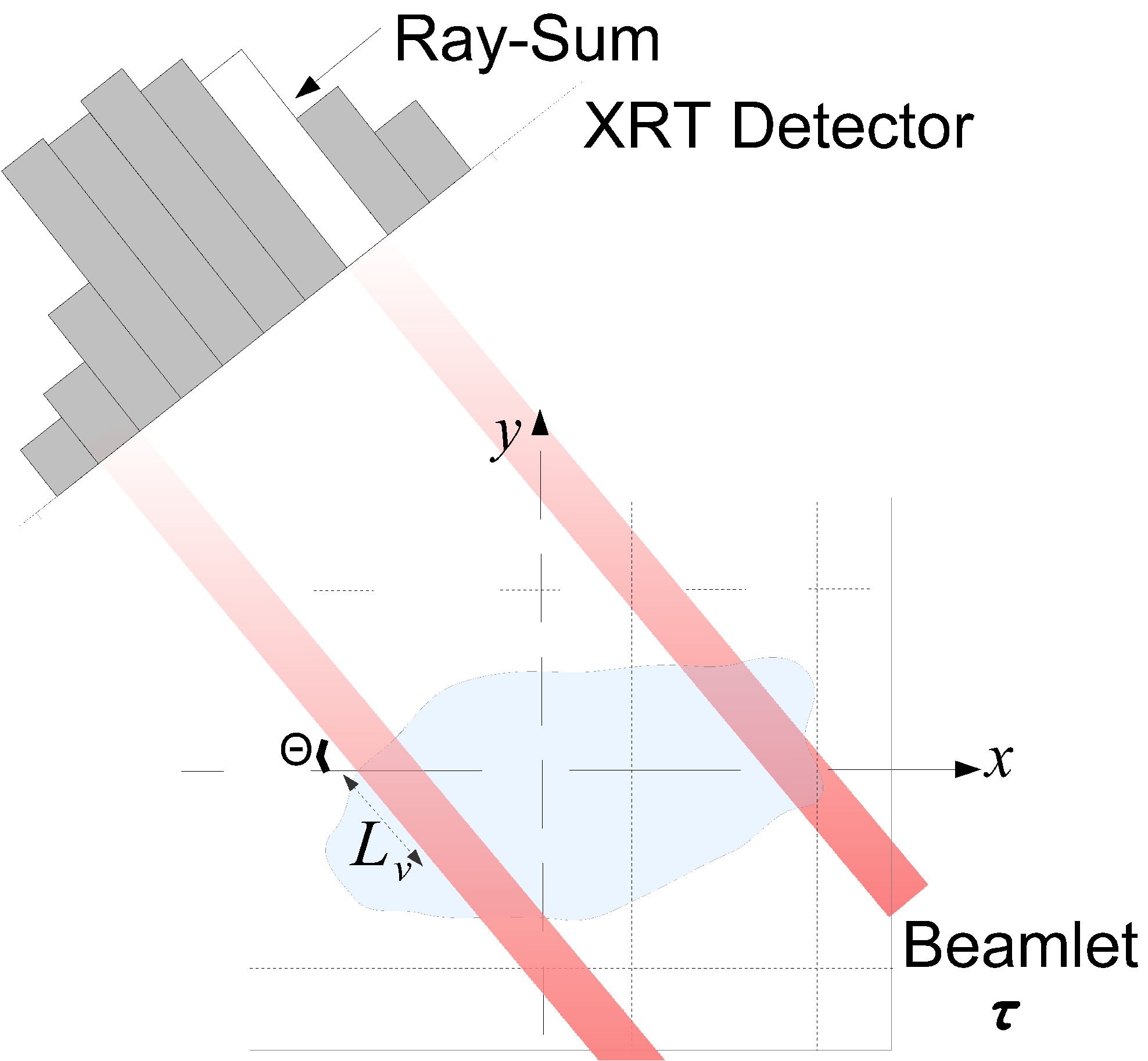}
\caption{Illustration of the discrete XRT projection geometry: the beam is parametrized by its angular and translation scan (indexed by $\theta$ and $\tau$, respectively).}
\label{discrete_xrt}
\end{figure}

In X-ray tomography, the Radon transform \cite{Radon-1986} provides the forward model from the object to the measurement. For simplicity, we focus on the 2D object in this work. Given a 2D discretized object $\bm{W}\in \R^{K \times K}$, $W_{ij}$ is the physical property of interest (i.e., the attenuation coefficient) at pixel $(i,j)$. $\Theta$ and $\T$ denote the complete collection of $\abs{\Theta}$ angles and $\abs{\T}$ beamlets, respectively. $\theta \in \Theta$ and $\tau \in \T$ denote the index of the X-ray beam angle and discretized beamlet, respectively. In tomographic imaging, the object is scanned with $\abs{\Theta}$ angles, and for each angle the X-ray beam is discretized into $\abs{\T}$ beamlets. Therefore, the whole scan results in a 2D sinogram (i.e., measurement data) $\bm{S} \in \R^{\abs{\Theta} \times \abs{\T}}$. Let the 3D tensor $\mathbb{L}= [L_{ij}^{\theta,\tau}] \in \R^{\abs{\Theta}\abs{\T} \times K \times K}$ represent the discrete Radon transform, where $L_{ij}^{\theta,\tau}$ denotes the intersection length of the beam $(\theta,\tau)$ with the pixel $(i,j)$. Then we have the final forward model mapping from the object space to the data space as $\bm{s} = \dotprod{\L,\bm{W}}_2$ , where $\bm{s} = \vect(\bm{S}) \in \R^{\abs{\Theta} \abs{\T} \times 1}$. Figure~\ref{discrete_xrt} illustrates the geometry of the discrete XRT projection \cite{Di-et-al-2016}. 

\section{Low Rank TENSOR REGRESSION} 

\label{Section3} 


In general, solving tomographic reconstruction results in a large-scale optimization problem due to its $K^2$ number of unknown parameters in 2D. 
To mitigate this issue, and inspired by \cite{Zhou-et-al-2013}, we explore a low-rank structure of $\bm{W}$ that admits a rank-$\tilde{R}$ CP decomposition: 
\begin{equation}\label{CP_decomp_W}
   \bm{W} = [\![\overset{\sim}{\bm{W}}_1,\overset{\sim}{\bm{W}}_2 ]\!] = \sum_{r=1}^{\tilde{R}} \bm{w}_1^{(r)} \circ \bm{w}_2^{(r)}.
\end{equation}
More important, instead of using its exact rank $\tilde{R}$, we further explore its low-rank approximation 
\begin{equation}\label{CP_decomp_W_approx}
   \bm{W} \approx [\![\bm{W}_1,\bm{W}_2 ]\!] = \sum_{r=1}^R \bm{w}_1^{(r)} \circ \bm{w}_2^{(r)},
\end{equation}
where $R\leq \tilde{R}$, $\bm{W}_1,\bm{W}_2\in \R^{K\times R}$, and $\bm{w}_1^{(r)},\bm{w}_2^{(r)}\in \R^{K\times 1}$. To formulate the final optimization problem, notice that similar to traditionally vectorized inverse problems, the maximum likelihood framework \cite{Shepp-Vardi-1982,rossi2018mathematical} can be applied to our proposed low-rank estimation as well, provided with the knowledge of the prior model distribution (i.e., Gaussian or Poisson). In this work, we focus on Gaussian distribution given the reality of high photon counts from tomographic imaging. Therefore, we convert the traditional vectorized linear least square problem to a tensor-based loss function $\phi$ as 
\begin{equation}
\label{eqn:obj_tensor}
\phi(\bm{W}_1,\bm{W}_2) =  \left \|\left \langle\mathbb{L},\sum_{r=1}^R \bm{w}_1^{(r)} \circ  \bm{w}_2^{(r)}\right \rangle_2-s\right\|^2.
\end{equation}
Essentially, the low-rank approximation tries to represent a high-dimensional tensor by combining smaller-dimensional sparse tensors. 
For example, to reconstruct a 2D discretized object with size $128\times 128$, compared with $128^2=16384$  unknown parameters arising from the vectorized least square model, our proposed low-rank reconstruction reduces the number of unknown parameters to $128 \times 2=256$ for a rank-1 model and $3\times 128\times 2=768$ for a rank-3 model. Such a massive reduction in dimensionality is shown to be a more efficient estimation \cite{Zhou-et-al-2013}. To solve the resulting decomposed components, we use a special type of block-relaxation algorithm, alternating least squares \cite{Kolda-Bader-2009}, to alternately update $\bm{W}_d, d =1 ,2$. 

 Tomographic reconstruction is often an ill-posed problem due to its limited data. It is critical to take advantage of any prior knowledge  either during the pre-/postprocessing stage or directly in the inverse problem framework \cite{verhoeven1993limited,oliveira2011comparison}.  For the latter, regularization techniques \cite{Tikhonov-Arsenin-1977,vauhkonen1998tikhonov} play an important role in penalizing any change of parameters violating the prior knowledge, which we exploit to solve problem~\ref{eqn:obj_tensor}. So, our final optimization problem is to minimize the following regularized least squares function:
\begin{equation*}
    l(\bm{W}_1,\bm{W}_2) = \phi(\bm{W}_1,\bm{W}_2) + \sum_{d=1}^2 \sum_{r=1}^R 
    P_{\lambda}(\bm{w}_{d}^{(r)},\rho),
\end{equation*}
where $P_{\lambda}$ is any regularization function, 
$\rho$ is the penalty parameter tuning the weight applied on the regularization, and $\lambda$ determines the weight from a specific penalty type. In particular, we apply  elastic net regularization \cite{Zou-Hastie-2005}, \[\ds P_{\lambda}(\bm{w},\rho) = \rho \left( \frac{\lambda-1}{2}\|\bm{w}\|_2^2 +(2-\lambda) \|\bm{w}\|_1\right ),\] where $\lambda \in [1,2]$, to simultaneously promote sparsity and smoothness through a convex combination of $L_1$ and $L_2$ penalties and improve the recovery of sharp as well as smooth features of the object. A complete description of our proposed low-rank tensor regression algorithm (denoted as TR($R$)($P_{\lambda}(\bm{w},\rho)$) is presented in Alg.~\ref{alg:TR}



\begin{algorithm}[H]
\caption{Low-rank tensor regression TR($R$)($P_{\lambda}(\bm{w},\rho)$).} \label{alg:TR}
\begin{algorithmic}[1]
\State Input: $s, \mathbb{L}, \bm{W}$, $R$, $\lambda$, $\rho$, maximum number of iterations $k_{\max}$ (e.g., 100) and stopping criterion $\epsilon$ (e.g., $10^{-4}$).
\State Initialize $\bm{W}_d^0 \in \R^{K\times R}$, for $d=1,2$. 
\For {$k= 1,2,\dots, k_{\max}$}
\State  $\bm{W}_1^k = \min\limits_{\bm{W}_1} \enskip l(\bm{W}_1,\bm{W}_{2}^{k-1})$
\State  $\bm{W}_2^k = \min\limits_{\bm{W}_2} \enskip l(\bm{W}_1^k,\bm{W}_2)$
\If {$\abs{l(\bm{W}_1^k,\bm{W}_2^k)-l(\bm{W}_1^{k-1},\bm{W}_2^{k-1})} < \epsilon$ }
\State \quad \textbf{break}
\EndIf
\EndFor
\State Output: construct $\bm{W}$ using \eqref{CP_decomp_W} from final $\bm{W}_1,\bm{W}_2$. 
\end{algorithmic}
\end{algorithm}

\section{Numerical Results} 

\label{Section4} 

\begin{figure}[htb]
\centering
\includegraphics[scale=0.5]{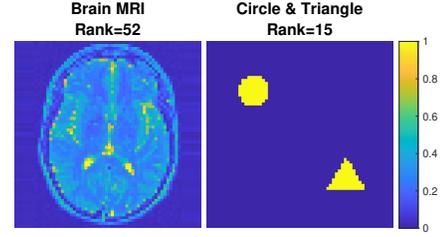}
\caption{Test images and colormap.}
\label{test_images}
\end{figure}

We have carried out extensive numerical analysis to investigate the performance of our proposed low-rank tensor regression algorithm. We choose two images as the ground truth samples (see Fig.~\ref{test_images}): a simple geometric shape consisting of circle and triangle and a real MRI brain image. For the experimental configuration, we fix the image resolution $K = 64$ and the number of discretized beamlets $\abs{\T} = 91> \sqrt{2}K$ to guarantee full coverage of the object from any angle. We adopt root mean squared error (RMSE) as our error metric to quantify the reconstruction quality. We perform most of the analyses with fixed $\abs{\Theta}=30$ angles in the range of $[1,2\pi]$ as well as varying numbers of angles to study the sensitivity of our proposed method towards the number of data.

\begin{figure}[htb]
\centering
\includegraphics[scale=0.5]{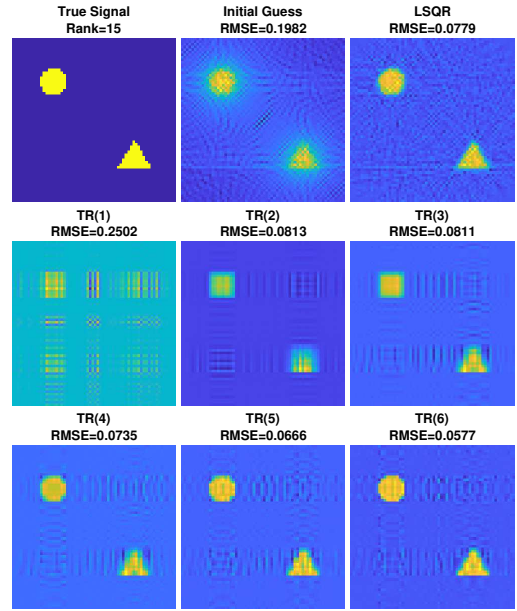}
\caption{Reconstructed image of the simple geometric shape using unregularized TR with $R=1,\dots,6$, and LSQR.}
\label{Circle_Triangle_Reconstruction_Image}
\end{figure}

To isolate the contribution of TR, we first demonstrate its performance on the simple geometric shape (Fig.~\ref{test_images} right) without applying any regularization. Notice that this ground truth image has a rank of 15, that is, $\ds \bm{W}= \sum\limits_{r=1}^{15} \bm{w}_{1}^{(r)} \circ \bm{w}_{2}^{(r)}$. In Fig.~\ref{Circle_Triangle_Reconstruction_Image}, we show the reconstruction results using different levels of low-rank approximations ranging from rank 1 to 6, and we compare their results with the  MATLAB built-in solver LSQR \cite{Paige-Saunders-1982},  a  popular iterative method for solving large linear systems of
equations and least-squares problems. We fix all the initial guesses to the backprojection solution \cite{Kak-Slaney-1988}, for example, $\bm{W}^0=\bm{L}_{(1)}^T s$, where $\bm{L}_{(1)}$ is the matrix obtained by unfolding $\mathbb{L}$ along the first dimension \cite{bader2006algorithm}.  



\begin{figure}[htb]
\centering
\includegraphics[scale=0.25]{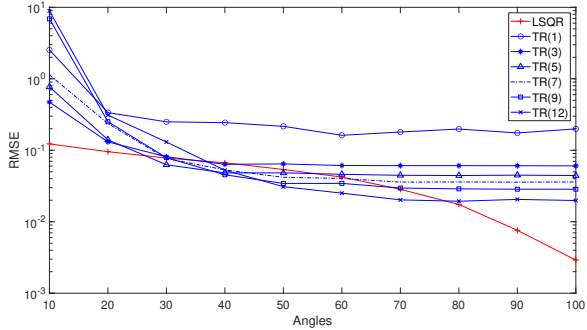}
\caption{Comparison of RMSE with varying numbers of angles between LSQR and unregularized TR.}
\label{Circle_triangle_unregularized_TR_lsqr_no_noise_rmse_plot}
\end{figure}

Next, we vary the number of angles from 10 to 100, which are evenly sampled within $[1,2\pi]$, to further illustrate the performance of TR. With low-rank approximations ranging from 1 to 12, Fig.~\ref{Circle_triangle_unregularized_TR_lsqr_no_noise_rmse_plot} shows the TR results compared with LSQR in terms of final RMSE. We observe that our proposed TR methods perform better than LSQR for most of the limited-angle cases. 

\begin{figure}[htb]
\centering
\includegraphics[scale=0.4]{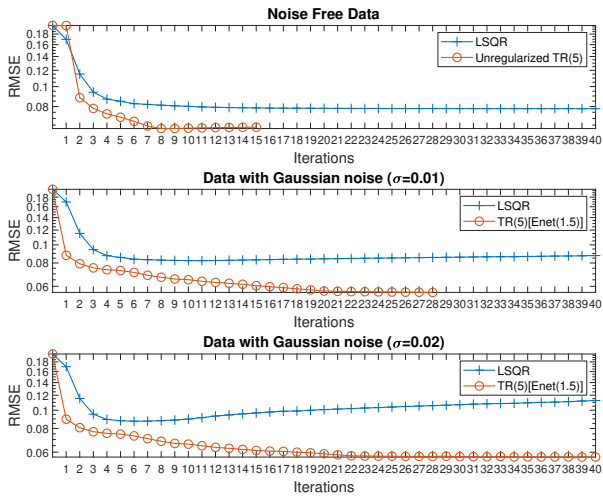}
\caption{Iterative performance of RMSE provided by LSQR and TR(5), respectively, for recovering the simple geometric shape. Top: noise-free data; middle: 1\% Gaussian noise-added data; bottom: 2\% Gaussian noise-added data.}
\label{Cir_Tri_rmse_iterations}
\end{figure}

Now we demonstrate the performance of TR with the elastic net regularizer for both the test images. For each case, we further check the robustness of the proposed method by adding 1\% and 2\% Gaussian noise to the data (relative to the maximum intensity of its corresponding noise-free data), respectively. For the geometric shape, given the result shown in Fig.~\ref{Circle_triangle_unregularized_TR_lsqr_no_noise_rmse_plot}, we fix the approximation rank $R=5$, $\lambda=1.5$, and $\rho=10^{-5}$ (trial and error) for the next experiment. The result is shown in Fig.~\ref{Cir_Tri_rmse_iterations} in terms of the convergence behavior of TR and LSQR given different noise levels. As we can see, with increasing noise levels, TR consistently outperforms LSQR in terms of both convergence speed and reconstruction quality. 



\begin{figure}[htb]
\centering
\includegraphics[scale=0.5]{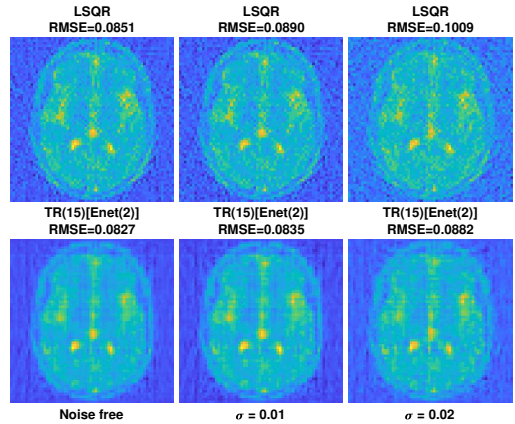}
\caption{Reconstruction comparison of TR and LSQR for the MRI brain image. Top: LSQR; bottom: TR(15)[Enet(2)]. Left: noise-free data; middle: 1\% Gaussian noise-added data; right: 2\% Gaussian noise-added data. }
\label{Brain_MRI_Reconstruction_Image}
\end{figure}

We also illustrate the performance of TR on the MRI brain image (Fig.~\ref{test_images} left). In this case, given that the full rank of the ground truth is 52, we choose the approximation rank $R=15$, $\lambda=2$, and $\rho = 10^{-5}$ (trial and error) for both the noise-free and Gaussian noise-added data. Figure~\ref{Brain_MRI_Reconstruction_Image} shows the reconstruction results from TR and LSQR with corresponding RMSE values. Again, we see that even with a relatively low rank approximation, TR outperforms LSQR in terms of reconstruction quality.

\section{CONCLUSION AND DISCUSSION} 

\label{Section5} 


In conclusion, we exploit the underlying structure of tomography to better capture its latent multilinear structure. Instead of solving the reconstruction problem as a standard linear least squares problem where the unknown parameters are vectorized, we explore the low-rank approximation of the unknown parameters in its natural tensor form (i.e., 2D matrix for an image) to mitigate the curse of dimensionality, as well as the ill-posed nature of tomography due to the limited data. For simplicity, we demonstrate the performance of our proposed tensor regression on a 2D reconstruction problem, where the numerical results show that the tensor method outperforms the traditional vectorized linear least square solver. Furthermore, our method is shown to be more robust to  limited number of angles and increasing levels of added noise. The extension of our proposed method to 3D reconstruction is natural, with potentially more dramatic benefit compared with traditional methods. 



\section*{Acknowledgments}
This material was based upon work supported by the U.S.\ Department of Energy under contract DE-AC02-06CH11357.

\clearpage


\bibliographystyle{IEEEbib}
\bibliography{ICIP21}

\end{document}